\documentclass[11pt]{article}%
\pdfoutput=1
\usepackage[hidelinks]{hyperref}
\usepackage{xurl,bbm,microtype,amssymb,amsthm,mathtools,mathrsfs,cleveref,graphicx,float,color}
\usepackage{algorithmic}
\usepackage{fullpage}
\usepackage{lmodern}
\usepackage[T1]{fontenc}
\usepackage{authblk}

\usepackage[USenglish]{babel}

\usepackage{longtable}
\usepackage{booktabs} 
\usepackage{caption}
\usepackage{subcaption}
\usepackage{multirow}

\usepackage{tikz} %
\usetikzlibrary{decorations.pathmorphing,decorations.pathreplacing,arrows.meta,positioning,fit,calc,backgrounds,shapes.arrows,matrix,intersections}
\usetikzlibrary{external}
\pgfdeclarelayer{background}
\pgfsetlayers{background,main}

\tikzset{%
  gate/.style={draw, minimum size=14, fill=gray!15, inner sep=2},
  control/.style={circle, fill=black, minimum size=3, inner sep=0},
  target/.style={circle, draw, minimum size=8, inner sep=0},
  cross/.style={cross out, draw, minimum size=2.5, inner sep=0},
  wire/classical/.style={double, double distance=1pt},
  wire/solid/.style={on background layer},
  wire/inbox/.style={dashed, line cap=round},
  cell/.style={rectangle, draw=black, minimum width=1cm, minimum height=1cm,
      text height=1.5ex, text depth=.25ex, font=\scriptsize, align=center},
  celllabel/.style={font=\scriptsize, align=center},
  labelsty/.style={font=\scriptsize, align=center},
  pics/tensorbox/.style n args={3}{
    code={
      \node[draw, minimum width=1cm, minimum height=2cm, fill=gray!15,
            rounded corners=2pt, align=center, font=\scriptsize, pic actions] (B) {#1};
      \foreach \i/\t in {0/0.11, 1/0.37, 2/0.63, 3/0.89}{
        \coordinate (-in\i) at ($(B.north west)!\t!(B.south west)$);
        \coordinate (-inpin\i) at ($(-in\i)+(-.2,0)$);
      }
      \draw[wire/classical] (-inpin0) -- (-in0);
      \node[font=\scriptsize, anchor=north, yshift=20pt, xshift=4pt] at (-inpin0) {#2};
      \foreach \i/\t in {0/0.11, 1/0.37, 2/0.63, 3/0.89}{
        \coordinate (-out\i) at ($(B.north east)!\t!(B.south east)$);
        \coordinate (-outpin\i) at ($(-out\i)+(.2,0)$);
      }
      \draw[wire/classical] (-out0) -- (-outpin0);
      \node[font=\scriptsize, anchor=north, yshift=20pt, xshift=2pt] at (-outpin0) {#3};
      \coordinate (-SW) at (B.south west);
      \coordinate (-NE) at (B.north east);
    }
  },
  pics/tensorboxin/.style n args={2}{
    code={
      \node[draw, minimum width=1cm, minimum height=1.5cm, fill=gray!15,
            rounded corners=2pt, align=center, font=\scriptsize, pic actions] (B) {#1};
      \foreach \i/\t in {0/0.11, 1/0.37, 2/0.63, 3/0.89}{
        \coordinate (-in\i) at ($(B.north west)!\t!(B.south west)$);
        \coordinate (-inpin\i) at ($(-in\i)+(-.2,0)$);
      }
      \draw[wire/classical] (-inpin0) -- (-in0);
      \node[font=\scriptsize, xshift=-8pt] at (-inpin0) {#2};
      \foreach \i/\t in {0/0.11, 1/0.37, 2/0.63, 3/0.89}{
        \coordinate (-out\i) at ($(B.north east)!\t!(B.south east)$);
        \coordinate (-outpin\i) at ($(-out\i)+(.2,0)$);
      }
    }
  },
  pics/tensorboxout/.style n args={2}{
    code={
      \node[draw, minimum width=1cm, minimum height=1.5cm, fill=gray!15,
            rounded corners=2pt, align=center, font=\scriptsize, pic actions] (B) {#1};
      \foreach \i/\t in {0/0.11, 1/0.37, 2/0.63, 3/0.89}{
        \coordinate (-in\i) at ($(B.north west)!\t!(B.south west)$);
        \coordinate (-inpin\i) at ($(-in\i)+(-.2,0)$);
      }
      \foreach \i/\t in {0/0.11, 1/0.37, 2/0.63, 3/0.89}{
        \coordinate (-out\i) at ($(B.north east)!\t!(B.south east)$);
        \coordinate (-outpin\i) at ($(-out\i)+(.2,0)$);
      }
      \draw[wire/classical] (-out0) -- (-outpin0);
      \node[font=\scriptsize, xshift=8pt] at (-outpin0) {#2};
    }
  },
  pics/tensorboxwide/.style n args={1}{
    code={
      \node[draw, minimum width=1.4cm, minimum height=1.8cm, fill=gray!15,
            rounded corners=2pt, align=center, font=\scriptsize, pic actions] (B) {#1};
      \foreach \i/\t in {0/0.11, 1/0.37, 2/0.63, 3/0.89}{
        \coordinate (-in\i) at ($(B.north west)!\t!(B.south west)$);
        \coordinate (-inpin\i) at ($(-in\i)+(-.2,0)$);
      }
      \foreach \i/\t in {0/0.11, 1/0.37, 2/0.63, 3/0.89}{
        \coordinate (-out\i) at ($(B.north east)!\t!(B.south east)$);
        \coordinate (-outpin\i) at ($(-out\i)+(.2,0)$);
      }
    }
  },
  pics/tensorboxmid/.style n args={1}{
    code={
      \node[draw, minimum width=1cm, minimum height=1.5cm, fill=gray!15,
            rounded corners=2pt, align=center, font=\scriptsize, pic actions] (B) {#1};
      \foreach \i/\t in {0/0.11, 1/0.37, 2/0.63, 3/0.89}{
        \coordinate (-in\i) at ($(B.north west)!\t!(B.south west)$);
        \coordinate (-inpin\i) at ($(-in\i)+(-.2,0)$);
      }
      \foreach \i/\t in {0/0.11, 1/0.37, 2/0.63, 3/0.89}{
        \coordinate (-out\i) at ($(B.north east)!\t!(B.south east)$);
        \coordinate (-outpin\i) at ($(-out\i)+(.2,0)$);
      }
    }
  }
}

\newcommand{\microspace}{\mspace{.5mu}} %
\newcommand{\bigket}[1]{\bigl\lvert\microspace#1%
  \microspace\bigr\rangle} %
\newcommand{\bigbra}[1]{\bigl\langle\microspace#1%
  \microspace\bigr\rvert} %

\providecommand{\U}[1]{\protect\rule{.1in}{.1in}}
\newtheorem{thm}{Theorem}\crefname{thm}{Theorem}{Theorems}
\crefname{lem}{Lemma}{Lemmas}
\crefname{prp}{Proposition}{Propositions}
\crefname{cor}{Corollary}{Corollaries}
\newtheorem{dfn}[thm]{Definition}\crefname{dfn}{Definition}{Definitions}
\crefname{section}{Section}{Sections}
\crefname{appendix}{Appendix}{Appendices}
\numberwithin{equation}{section}
\let\oldref\ref
\renewcommand{\ref}[1]{(\oldref{#1})}

\newcommand{\ket}[1]{\vert #1\rangle }

\AddToHook{cmd/bfseries/after}{\boldmath}
\AddToHook{cmd/normalfont/after}{\unboldmath}



\begin{document}
\title{Quantum Telepathy: A Quantum Technology \\with Near-Term Applications}
\author[1,2]{Dawei Ding\thanks{daweiding@fudan.edu.cn}}
\author[2,3]{Xinyu Xu}
\affil[1]{\small Center for Mathematics and Interdisciplinary Sciences, Fudan University, Shanghai 200433, China}
\affil[2]{\small Shanghai Institute for Mathematics and Interdisciplinary Sciences (SIMIS), Shanghai 200433, China}
\affil[3]{\small Research Institute of Intelligent Complex Systems, Fudan University, Shanghai 200433, China}
\date{}
\maketitle

\begin{abstract}
    Quantum telepathy is the concept of using quantum entanglement to solve real-world problems involving decision coordination between parties with restricted communication. One possible reason for this restriction is a latency constraint: some pairs of parties do not have enough time to communicate with each other before they have to produce their outputs. Example scenarios include high frequency trading and distributed systems. Another reason is physical or operational isolation: for some pairs of parties, there is an obstacle to communication. Example scenarios include locating a stray traveler by a rescue team and coordination within a network where nodes are owned by competing firms. In this paper we give a concise overview of the different application areas of quantum telepathy. We find that these real-world problems can be modeled as a nonlocal game or its generalizations. We also discuss possible physical implementations. Quantum telepathy guarantees a quantum advantage via Bell's theorem and can directly solve real-world problems, such as reducing risk in high frequency trading or balancing data loads efficiently in ad hoc networks. Moreover, this quantum advantage can be physically realized with existing or near-term quantum hardware. 
\end{abstract}

\section{Introduction}
The holy grail of quantum computing and quantum information is to identify a quantum technology with practical applications and to physically build it. For decades this has been the dream of the scientists and engineers that have poured their time, energy, PhD students, and research funding into this quest. There has been remarkable progress: we now have quantum computers on the order of thousands of qubits~\cite{castelvecchi2023ibm}, quantum networks that span thousands of kilometers~\cite{yin2017satellite}, and error-corrected logical qubits below the fault-tolerant threshold~\cite{google2025quantum}. Despite these awesome feats, there is still an elephant in the room: what are the practical applications of the technologies we have developed so far? Yes, we can factorize integers to break RSA-2048 using Shor's algorithm with a million physical qubits~\cite{gidney2025factor}, but we do not have such hardware capabilities at the moment. In fact, we are still lacking by orders of magnitude. But surely, 
there should be a practical application within reach of \emph{current or near-term} quantum devices?

In this paper, we provide a concise overview of results concerning a new type of quantum technology that is a candidate for realizing the long-anticipated applications of near-term quantum devices. This potential has been highlighted in review articles and papers discussing near-term applications of quantum technologies~\cite{huang2025vast,huang2025comprehensive,beauchamp2025white,beauchamp2025towards,gauthier2025arqon}, as well as various media outlets~\cite{newscientist,quantuminsider1,quantuminsider2,quantuminsider3,crn}. The new technology takes advantage of quantum nonlocality: using quantum entanglement to produce correlations between parties with restricted communication that are impossible to achieve classically. Hence, we propose the term \textbf{quantum telepathy}~\cite{ding2024coordinating} for this technology. We emphasize that ``telepathy'' here does not refer to superluminal communication, but rather to the uncanny phenomenon\footnote{In Einstein's words, ``spooky action at a distance.'' } in which multiple parties can generate \emph{superclassical} correlations using quantum entanglement. Furthermore, our use of the term should be distinguished from ``quantum pseudotelepathy,'' a term used in previous literature~\cite{brassard2005quantum} to refer to the achievement of unit success probability in a nonlocal game using quantum resources.

One example of quantum nonlocality is Bell inequality violation. A Bell inequality is a bound on the correlations between multiple parties that are not in communication and can only use classical resources. However, as proved by John Bell himself~\cite{bell1964einstein} and by subsequent experimental verification~\cite{freedman1972experimental,fry1976experimental,aspect1982locality,weihs1998violation}, quantum mechanics can violate these inequalities. Another example of quantum nonlocality is the generalization of Bell inequalities to the case where \emph{a subset of parties can communicate}. This is described by the framework of latency-constrained games~\cite{ding2025quantum}, which we will describe more in~\Cref{sec:lc}. A multi-round extension of Bell inequality violation~\cite{da2025entanglement,gardiner2026learning} and quantum advantages in communication complexity~\cite{brassard2003quantum} are further manifestations of quantum nonlocality.

\emph{Quantum telepathy is the idea that we can use quantum nonlocality to address real-world problems}. But in our current information age with widespread internet access and near-global cellular networks, when is communication ever restricted? We list here some examples:
\begin{itemize}
    \item The parties are taking actions so fast that some cannot communicate due to communication latency. If the communication latency is the speed of light delay, non-communication is due to fundamental physical limitations imposed by the theory of relativity. Such situations can arise in high frequency trading (HFT)~\cite{brandenburger2016team,szegedy2020systems,ding2024coordinating,ding2025quantum}, where trades are executed at microsecond timescales --- during which light can only travel a few hundred meters in vacuum. Hence, if we want to realize correlations between trading servers located at different stock exchanges at these timescales for hedging~\cite{ding2024coordinating} or order routing~\cite{brandenburger2016team}, quantum telepathy may be a solution. A similar logic applies to distributed systems where multiple servers are running computations at GHz clock speeds while separated by distances of a hundred meters (square root of the typical area of a data center~\cite{techjury}) or greater. The servers may need to coordinate decisions, such as routing data, handling customer requests, or performing memory read operations~\cite{brandenburger2015quantum,hasanpour2017quantum,da2025entanglement,arun2025faster,ding2024coordinating,gardiner2026learning}.  We will call this type of real-world scenario a \textbf{latency-constrained scenario. } 
    \item There is an obstacle to communication for some of the parties. This could be a physical barrier between the parties, a lack of means to communicate, or a privacy concern~\cite{brandenburger2015quantum,szegedy2020systems,mironowicz2023entangled}. For example, a scientific expedition exploring the deep ocean deploys a team of drones to map out an underwater cave. Due to the thick cave walls and the limited battery supply of the drones, they cannot communicate with each other. However, they might wish to coordinate their decisions, such as rendezvousing at one of multiple designated checkpoints~\cite{mironowicz2023entangled}. We will call this type of real-world scenario an \textbf{isolated-party scenario}. 
\end{itemize}
\noindent Now, if the parties' communication is restricted due to these or other reasons, we can mathematically prove that their correlations must satisfy Bell inequalities~\cite{bell1964einstein}, or generalizations thereof~\cite{ding2025quantum,brassard2003quantum}. This limits how well classical parties can coordinate their decisions. However, such inequalities can be violated using quantum entanglement. In other words, \emph{the parties can achieve better coordination (better hedging, better routing, better rendezvous) in these real-world scenarios using quantum entanglement. } This is a mathematically provable quantum advantage that does not require any complexity-theoretic assumptions such as $\mathrm{BQP}\neq \mathrm{BPP}$. In fact, the proof can be very simple and succinct, such as in the case of the Clauser-Horne-Shimony-Holt (CHSH) inequality~\cite{CHSH}. Indeed, the initial motivation for Bell inequalities was to find a separation between quantum and classical mechanics ---  specifically, to identify phenomena that are possible quantum mechanically but are impossible classically. This is precisely the definition of a quantum advantage.

Furthermore, \emph{our current hardware capabilities are sufficient to realize this quantum advantage}. In fact, we were already able to realize it half a century ago with early experiments that violated the CHSH inequality~\cite{freedman1972experimental,fry1976experimental,aspect1982locality,weihs1998violation}. These and other experiments eventually led to the 2022 Nobel Prize in physics~\cite{aspect2022nobel}. Today, Bell inequality experiments have already advanced to the point where the two parties can be thousands of kilometers apart~\cite{yin2017satellite}. Furthermore, the speed of the measurements is now faster than the speed of light delay, and simultaneously, the efficiency of the photon detectors is now high enough that the quantum advantage is statistically significant~\cite{hensen2015loophole,shalm2015strong,giustina2015significant}. In~\cite{ding2024coordinating}, a back-of-the-envelope calculation showed that our current hardware capabilities can already support quantum-enhanced HFT between a trading server at the New York Stock Exchange (NYSE) and a trading server at NASDAQ dozens of kilometers away. The Cisco Research quantum networks team also emphasized that the hardware for quantum telepathy is already available~\cite{ciscosummit}.
Indeed, hardware-intensive protocols, such as fault-tolerant quantum computing, are not always necessary to realize a quantum advantage. For example, violating the CHSH inequality simply needs two physical qubits and single-qubit operations. 
We also mention that recent breakthroughs in optical interfacing for neutral atom arrays~\cite{Shaw2026CavityArray} have made it possible to quickly scale up quantum memories, a technological advance that can make quantum telepathy much more practical. 
In fact, a concrete architecture for quantum telepathy applied to HFT and other real-world settings using an array of neutral atom quantum memories coupled to optical cavities was presented in~\cite{Li2026LCTC}. 

Lastly, there is a fundamental argument for why quantum telepathy is much more sensible from a hardware implementation perspective, compared to quantum computing for instance. To achieve a quantum advantage, the objective in quantum computing is to run a quantum algorithm to solve a classically hard problem. The goal is to obtain a specific quantum state or a specific measurement output distribution. There is some tolerance for error, but the \emph{objective} is defined by that specific state or distribution. In contrast, the objective in quantum telepathy is to violate a Bell inequality. In particular, we do not need to obtain a specific value of the violation ---  we only need to obtain a nonzero violation. A larger violation is of course preferable, but not necessary to achieve a quantum advantage.
\emph{Hence, the objective of quantum computing is inherently sensitive to noise, while the objective of quantum telepathy is inherently robust to noise}. This argument is similar to that of the digital abstraction~\cite{shannon1938symbolic}, the key conceptual leap that made classical computing a viable technology: an analog voltage \emph{value} is inherently sensitive to noise, while a digital bit (0 or 1) defined by a voltage \emph{threshold} is inherently robust.
These facts together constitute a compelling argument that quantum telepathy is indeed a promising quantum technology with near-term applications.

\section{Crash Course on Nonlocal Games}
\label{sec:prelim}
In this section we first briefly introduce some of the mathematical language behind quantum telepathy.
It is convenient to use the mathematical framework of \textbf{nonlocal games} from computer science. However, we will make some small changes in terminology in order to make explicit the connection to real-world applications. We will also try to make the concepts more concrete by using HFT as an example.  Most technical details can be skipped without affecting comprehension of the rest of the paper. 

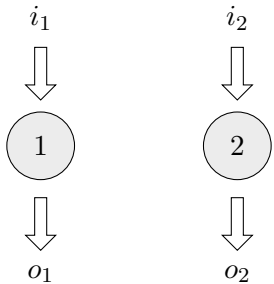
\begin{figure}[htbp!]
    \centering
    \begin{tikzpicture}[node distance=1.7cm,
  player/.style = {draw, circle, minimum width=.9cm, fill=gray!15},
  singlearrow/.style = {single arrow, draw, inner sep=0pt,
    minimum height=.7cm, minimum width=4mm, single arrow head extend=.3pt, rotate=-90}]
  
  \node[player] (P1) at (0,0) {$1$};
  \node[player, right=of P1] (P2) {$2$};        


  \node[singlearrow] at ([yshift=1cm]P1) {};
  \node[singlearrow] at ([yshift=1cm]P2) {};
  \node at ([yshift=1.7cm]P1) {$i_{1}$};
  \node at ([yshift=1.7cm]P2) {$i_{2}$};

  \node[singlearrow] at ([yshift=-1cm]P1) {};
  \node[singlearrow] at ([yshift=-1cm]P2) {};
  \node at ([yshift=-1.7cm]P1) {$o_{1}$};
  \node at ([yshift=-1.7cm]P2) {$o_{2}$};
\end{tikzpicture}
\caption{A nonlocal game with two parties. Each party $j$ receives input $i_j$ and produces output $o_j$. The parties cannot communicate throughout this process. }
\label{fig:nonlocal_game}
\end{figure}

A nonlocal game involves multiple parties who are receiving inputs and producing outputs, but cannot communicate with each other. In HFT, these parties could be colocated servers at different stock exchanges. In this section we will only give definitions for two-party nonlocal games. The definitions for more than two parties are straightforward generalizations.
The setup is shown diagrammatically in~\Cref{fig:nonlocal_game}.

\noindent We give a formal definition:
\begin{dfn}
    Let $I_1,I_2,O_1,O_2$ be finite sets. $I_1,I_2$ are \textbf{input sets}, and $O_1, O_2$ are \textbf{output sets}. We define a \textbf{utility function} $\mathcal{U}$ which is a map that assigns a real number to the outputs given the inputs:
    \begin{align*}
        \mathcal U: O_1 \times O_2  \times  I_1 \times I_2 \to \mathbb R.
    \end{align*}
    Furthermore, we define the \textbf{input distribution} $\pi$ which is a probability distribution on $I_1 \times I_2$. A \textbf{nonlocal game} is described by the tuple $(\mathcal{U},\pi)$.
\end{dfn}
\noindent This definition is to be interpreted as follows. Each party $j$ receives an input from the set $I_j$ and produces an output from the set $O_j$. In HFT, $I_j$ could be the set of all possible real-time local exchange information such as stock prices, and $O_j$ could be the set of all possible trade decisions, such as buying or selling a specific stock. 
$\pi$ is the probability distribution over the possible inputs and the utility function $\mathcal{U}$ is the ``score'' the parties receive for their outputs, given their inputs. These could respectively correspond to the prior distribution of the stock price movements on the different exchanges and the payoff of the servers' trade decisions given the exchanges' stock prices.
The parties' goal is to maximize the utility they achieve averaged over the input distribution $\pi$. To this end they can agree on a strategy in advance, but once the game starts, communication is forbidden. This makes sense from an HFT standpoint: the servers can formulate a strategy before the trading day begins, but once trading starts, due to the speed of light delay, they cannot communicate local exchange information in real time.

Implementation of the formulated strategy realizes a \textbf{behavior}, which is a probability distribution of the parties' joint outputs given their joint inputs. This corresponds to the distribution over the servers' joint trade decisions given the local exchanges' stock prices.
It is defined as a conditional probability distribution
$$ p(o_1, o_2 \vert i_1, i_2) \text{ where } i_j \in I_j, o_j \in I_j.$$
By combining the behavior $p$ with the utility function $\mathcal U$  and the input distribution $\pi$, we can compute the \textbf{average utility}:
\begin{align*}
    \bar u \coloneqq \sum_{i_j \in I_j, o_j \in O_j} \pi(i_1, i_2) p(o_1, o_2\vert i_1, i_2)  \mathcal U(o_1,o_2 , i_1, i_2).
\end{align*}
This could correspond to the average return on trades in HFT.

We next define what strategies the parties can implement with only classical resources. Given the parties cannot communicate, the most natural strategy is to give up on coordination and let each party produce his output based only on his own input. This leads to \textbf{deterministic strategies}:
\begin{dfn}
\label{dfn:classical}
    Let $(\mathcal{U},\pi)$ be a nonlocal game. A \textbf{deterministic strategy} is given by the functions ${\{f_1, f_2\}}$, where 
    $$f_j:I_j\rightarrow O_j.$$
    The behavior realized by a deterministic strategy is where each party $j$, when receiving input $i_j$, produces the output $f_j(i_j)$:
    \begin{align*}
    p(o_1, o_2\vert i_1, i_2)=\delta_{o_1,f_1(i_1)}\delta_{o_2,f_2(i_2)},
    \end{align*}
    where $\delta$ is the Kronecker delta function.
\end{dfn}
\noindent Such strategies are exactly what is currently being used in HFT: each server only sees their own local exchange information and accordingly makes a high-frequency trade. This is because coordination at microsecond timescales is assumed to be impossible in the current paradigm.
More generally, for a \textbf{classical strategy}, the parties can also make use of shared randomness. A classical strategy can thus be interpreted as a probabilistic mixture of deterministic strategies. The behavior realized is therefore a convex combination of deterministic behaviors. A \textbf{Bell inequality} can then be defined as the average utility achieved by classical strategies being upper bounded by the maximum attainable value:
$$ \bar u \leq c^* \coloneqq \max_\text{classical strategies} \bar u.$$
We refer to $c^*$ as the \textbf{classical value}. In HFT, a Bell inequality can correspond to the maximum average return if the trading servers can only use classical resources.

In contrast, in a \textbf{quantum strategy}, the parties have access to a shared entangled state and perform measurements dependent on the inputs they each receive:
\begin{dfn}
\label{dfn:quantum}
    Let $(\mathcal{U},\pi)$ be a nonlocal game. Let $B_1,B_2$ be finite-dimensional Hilbert spaces. A \textbf{quantum strategy} is a tuple $(\ket{\psi},{\{M_1(i_1), M_2(i_2)\}}_{i_1 \in I_1, i_2 \in I_2})$, where 
    $$\ket{\psi}\in B_1 \otimes B_2$$
    is a quantum state and 
    $$M_j(i_j)={\{\Pi_{j, o_j}(i_j)\}}_{o_j\in O_j}$$
    is a projective measurement on $B_j$ for each $i_j$. A quantum strategy realizes the behavior
    \begin{align*}
    p(o_1, o_2\vert i_1, i_2)=\bigbra{\psi}\Pi_{1, o_1}(i_1) \otimes \Pi_{2,o_2}(i_2) \bigket{\psi}.
\end{align*}
\end{dfn}
\noindent 
It is easy to see that classical behaviors are a subset of quantum behaviors. In fact, the quantum set is usually strictly larger, and this opens the possibility of achieving a higher average utility than what is classically possible for some nonlocal games:
$$ q^* \coloneqq \sup_\text{quantum strategies} \bar u >  c^*.$$
This phenomenon is called \textbf{Bell inequality violation}, and we call $q^*$ the \textbf{quantum value}. Equivalently, we say that there is a \textbf{quantum advantage} for this nonlocal game.
There are well-known nonlocal games for which this occurs, including the CHSH game~\cite{CHSH}, the magic square game~\cite{arkhipov2012extending}, and the GHZ game~\cite{greenberger1989going}. \emph{It is exactly this phenomenon, and its generalizations, that forms the theoretical basis for quantum telepathy's quantum advantage}. For example, in HFT this can be interpreted as achieving a provably higher average return on trades using quantum resources than what is possible with classical resources.

\section{Latency-Constrained Scenarios}
\label{sec:lc}
\begin{figure}[htbp!]
    \centering
    \begin{tikzpicture}[node distance=1.7cm,
  player/.style = {draw, circle, minimum width=.9cm, fill=gray!15},
  singlearrow/.style = {single arrow, draw, inner sep=0pt,
    minimum height=.7cm, minimum width=4mm, single arrow head extend=.3pt, rotate=-90}]
  
  \node[player] (P1) at (0,0) {$1$};
  \node[player] (P2) at ($(P1) +(4.4,0)$) {$2$};       

\draw[<->] (P1) -- node[midway, fill=white] {$d$} (P2);

  \node[singlearrow] at ([yshift=1cm]P1) {};
  \node[singlearrow] at ([yshift=1cm]P2) {};
  \node at ([yshift=1.7cm]P1) {$i_{1}$ at $t=0$};
  \node at ([yshift=1.7cm]P2) {$i_{2}$ at $t=0$};

  \node[singlearrow] at ([yshift=-1cm]P1) {};
  \node[singlearrow] at ([yshift=-1cm]P2) {};
  \node at ([yshift=-1.7cm]P1) {$o_{1}$ at $t < d/c$};
  \node at ([yshift=-1.7cm]P2) {$o_{2}$ at $t < d/c$};
\end{tikzpicture}
\caption{A latency-constrained scenario where non-communication is enforced by relativity. The parties have to produce their outputs after they receive their inputs within a time window shorter than their speed of light delay. Here, $c$ is the speed of light in vacuum. }
\label{fig:bell_latency}
\end{figure}
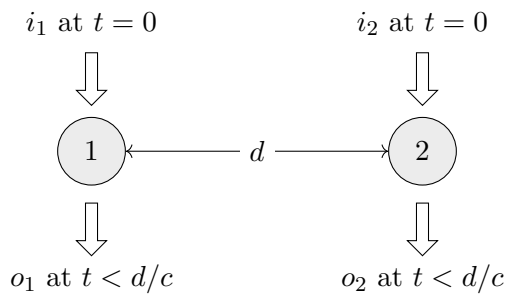

We now consider specific real-world scenarios that can be modeled by nonlocal games. As mentioned above, the most distinctive feature of a nonlocal game is that the parties cannot communicate.
One natural reason for this restriction is a \emph{latency constraint}. That is, after the parties receive their inputs, they must produce their outputs within a time window so short that it is impossible to communicate. In fact, if the time window is shorter than the speed of light delay between different parties, then \emph{non-communication is enforced by fundamental physics}, in particular by the theory of relativity. 
We show schematically this type of latency-constrained scenario in~\Cref{fig:bell_latency}.

\noindent This observation actually leads to an intuitive physical interpretation of Bell inequalities:
\begin{center}
    ``Bell inequalities are fundamental limits on the correlations between multiple classical parties \\
    acting within time windows shorter than the speed of light delay between any pair of parties.''
\end{center}
This interpretation is usually not how Bell inequalities are taught in the classroom, where a course typically starts with a mathematical definition of local hidden variable theories. However, this interpretation is prevalent in experimental papers that discuss the locality loophole~\cite{aspect1982locality,weihs1998violation}: measurements on particles must be performed within the speed of light delay between the detectors after the measurement settings are determined. In fact, it may be valuable to take this interpretation \emph{as the starting point} for Bell inequalities, as the latency constraint physically enforces non-communication.
Note however that for most real-world scenarios the communication latency can be longer than the speed of light delay. Possible reasons include light needing to travel through a medium such as optical fiber or a lack of a straight-line connection between the parties. Hence, non-communication can persist in these scenarios even for longer timescales.

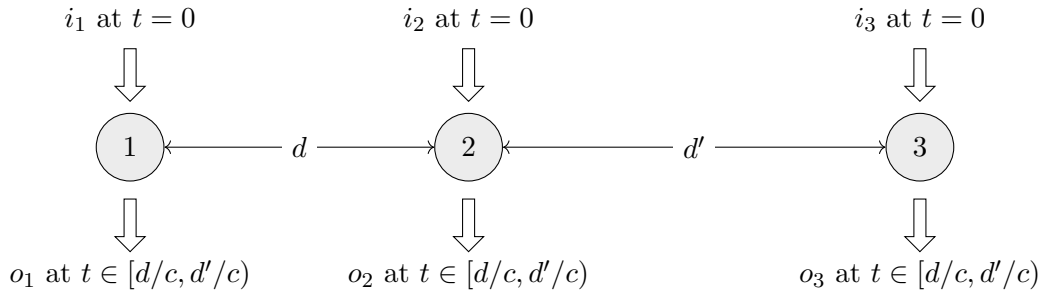
\begin{figure}[htbp!]
    \centering
    \begin{tikzpicture}[node distance=1.7cm,
  player/.style = {draw, circle, minimum width=.9cm, fill=gray!15},
  singlearrow/.style = {single arrow, draw, inner sep=0pt,
    minimum height=.7cm, minimum width=4mm, single arrow head extend=.3pt, rotate=-90}]
  
  \node[player] (P1) at (0,0) {$1$};
  \node[player] (P2) at ($(P1)+(4.5,0)$) {$2$}; 
  \node[player] (Pn) at ($(P2)+(6,0)$) {$3$}; 

  \draw[<->] (P1) -- node[midway, fill=white] {$d$} (P2);
  \draw[<->] (P2) -- node[midway, fill=white] {$d'$} (Pn);

  \node[singlearrow] at ([yshift=1cm]P1) {};
  \node[singlearrow] at ([yshift=1cm]P2) {};
  \node[singlearrow] at ([yshift=1cm]Pn) {};
  \node at ([yshift=1.7cm]P1) {$i_{1}$ at $t=0$};
  \node at ([yshift=1.7cm]P2) {$i_{2}$ at $t=0$};
  \node at ([yshift=1.7cm]Pn) {$i_{3}$ at $t=0$};

  \node[singlearrow] at ([yshift=-1cm]P1) {};
  \node[singlearrow] at ([yshift=-1cm]P2) {};
  \node[singlearrow] at ([yshift=-1cm]Pn) {};
  \node at ([yshift=-1.7cm]P1) {$o_{1}$ at $t \in [d/c, d'/c)$};
  \node at ([yshift=-1.7cm]P2) {$o_{2}$ at $t \in [d/c, d'/c)$};
  \node at ([yshift=-1.7cm]Pn) {$o_{3}$ at $t \in [d/c, d'/c)$};
\end{tikzpicture}
\caption{A latency-constrained scenario with three parties. The parties have to produce an output after they receive their inputs within a time window longer than the speed of light delay between the left two parties but shorter than that of the right two parties. This allows for communication only between the left two parties. }
\label{fig:partial_comm}
\end{figure}

In~\cite{ding2025quantum}, it was realized that if we can interpret Bell inequalities as arising from latency constraints, similar inequalities can be obtained by relaxing the latency constraints such that
\emph{a subset of parties can communicate}.\footnote{We are assuming we have more than two parties. } One such latency-constrained scenario with three parties is shown schematically in~\Cref{fig:partial_comm}. 

\noindent The paper then introduced the mathematical framework of \textbf{latency-constrained (LC) games} describing such scenarios~\cite{ding2025quantum}.
Using this framework, we can derive inequalities on correlations between classical parties where a subset of parties can communicate, leading to a very natural generalization of Bell inequalities which we call \textbf{LC inequalities}. This generalization, in turn, clarifies the nature of Bell inequalities as limitations on correlations arising from latency constraints.
LC inequalities can, just like Bell inequalities, be violated if the parties can use quantum resources. 
Note that since the latency constraint allows for a subset of parties to communicate, if they can use quantum resources, the parties can send quantum information to each other. That is, allowing partial communication leads to corresponding modifications of the definitions of classical and quantum strategies in Definition~\ref{dfn:classical} and Definition~\ref{dfn:quantum}, respectively. Detailed definitions can be found in~\cite{ding2025quantum}. In~\cite{ding2025quantum}, multi-round latency-constrained scenarios are also considered where the parties are receiving multiple inputs and producing multiple outputs over the course of time. Inputs can be communicated with each other, but with a latency cost. In other words, inputs and other information communicated can only affect the outputs that are produced at a later time. We can again derive corresponding inequalities on the correlations between classical parties in such a scenario. 

The real-world scenarios we analyze in this paper are modeled by nonlocal games where no parties can communicate. However, the same scenarios yield LC games if we increase the number of parties and allow a subset of parties to communicate. A quantum advantage can still exist after such a generalization. We could also modify the scenario by allowing for multiple rounds of inputs and outputs. 

\subsection{High frequency trading}
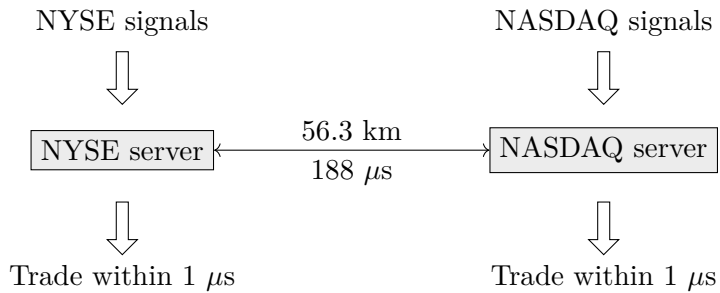
\begin{figure}[htbp!]
    \centering
    \begin{tikzpicture}[node distance=1.7cm,
  player/.style = {draw, rectangle, minimum width=.9cm, fill=gray!15},
  singlearrow/.style = {single arrow, draw, inner sep=0pt,
    minimum height=.7cm, minimum width=4mm, single arrow head extend=.3pt, rotate=-90}]
  
  \node[player] (P1) at (0,0) {NYSE server};
  \node[player] (P2) at ($(P1) +(6.4,0)$) {NASDAQ server};       

\draw[<->] (P1) -- node[midway, above] {$56.3$ km} node[midway, below] {188 $\mu$s} (P2);

  \node[singlearrow] at ([yshift=1cm]P1) {};
  \node[singlearrow] at ([yshift=1cm]P2) {};
  \node at ([yshift=1.7cm]P1) {NYSE signals};
  \node at ([yshift=1.7cm]P2) {NASDAQ signals};

  \node[singlearrow] at ([yshift=-1cm]P1) {};
  \node[singlearrow] at ([yshift=-1cm]P2) {};
  \node at ([yshift=-1.7cm]P1) {Trade within 1 $\mu$s};
  \node at ([yshift=-1.7cm]P2) {Trade within 1 $\mu$s};
\end{tikzpicture}
\caption{A latency-constrained scenario involving two colocated servers engaged in high frequency trading at NYSE and NASDAQ. Trades can be executed within 1 $\mu$s or shorter of a signal, whereas the speed of light delay between the two servers is 188 $\mu$s. }
\label{fig:hft}
\end{figure}

Latency-constrained scenarios can often appear in high frequency trading (HFT)~\cite{brandenburger2016team,szegedy2020systems,ding2024coordinating}. In such scenarios, servers engaged in HFT are physically located in data centers of different stock exchanges. We call such servers \textbf{colocated}. Today, trades can be executed by servers within microseconds (or shorter) of a price update or other signal from the exchange~\cite{haldane2010}. In comparison, the speed of light delay between exchanges can be much longer. For example,~\Cref{fig:hft} shows colocated servers at NYSE and NASDAQ, whose data centers are separated by 56.3 km. This translates to a speed of light delay of 188 $\mu$s, much longer than the timescale of a high frequency trade.

Now, these servers can engage in HFT independently very well using established technologies. The problem arises when we wish to \emph{coordinate high frequency trades}. That is, given the signals observed at \emph{both} exchanges, we wish to make a corresponding pair of trades within a very short time window. When this time window is shorter than the speed of light delay, a latency-constrained scenario arises where communication between the servers becomes physically impossible. We can then model the scenario by a nonlocal game. The input $i_j$ is the signal received by server $j$, the output $o_j$ is the trade executed, and the utility function $\mathcal U$ is a global ``score'' given to the pair of trades executed given the signals at both exchanges. This score could be payoff, risk,\footnote{Or rather negative risk since we are trying to maximize the utility. } or some other figure of merit. Lastly, the input distribution $\pi$ is a prior over possible signals the servers can receive. This could be determined using historical data, for example.

In~\cite{ding2024coordinating}, a concrete trading scenario was proposed which can be modeled by the CHSH game. Specifically,\footnote{Note that this trading scenario is considerably simplified and should be interpreted as a toy model. } consider the setting shown in~\Cref{fig:hft}. Suppose the NYSE server trades stock X and the NASDAQ server trades stock Y. The servers issue buy or sell orders for the corresponding stock. However, they wish to do this in a coordinated fashion in order to minimize risk. In particular, stock X and Y are correlated, but the sign of their correlation (positive or negative) can change over time. The two servers each look for a signal that shows the correlation has flipped. Suppose that X and Y are positively correlated. Then, in order to minimize risk, the two servers should issue opposite orders (one buy and one sell). By the same logic, when X and Y are negatively correlated, the servers should issue the same orders (both buy or both sell). 

Now, assume stock $X$ and stock $Y$ are initially negatively correlated. If we assume both servers need to look for this signal to conclude that the stocks' correlation has indeed flipped and that the signals are uniform, this trading scenario is exactly modeled by the CHSH game:
\begin{align*}
    p_\text{success} &= \frac 1 4 [ p(o_1 =o_2 \vert i_1 =0, i_2 =0) + p(o_1 = o_2 \vert i_1 =0, i_2 = 1) + p(o_1 = o_2 \vert i_1 = 1, i_2 = 0) \\
    & + p(o_1 \neq o_2 \vert i_1 =1, i_2 = 1) ].
\end{align*}

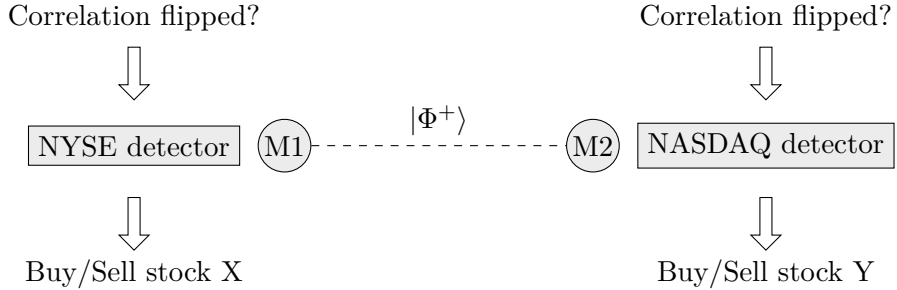
\begin{figure}[htbp!]
    \centering
    \begin{tikzpicture}[node distance=1.7cm,
  player/.style = {draw, rectangle, minimum width=.9cm, fill=gray!15},
  particle/.style = {draw, circle, minimum size=.7cm, inner sep = 0.3, fill=gray!15},
  singlearrow/.style = {single arrow, draw, inner sep=0pt,
    minimum height=.7cm, minimum width=4mm, single arrow head extend=.3pt, rotate=-90}]
  
  \node[player] (P1) at (0,0) {NYSE detector};
  \node[particle] (E1) at ($(P1)+(2,0)$) {M1};
  \node[player] (P2) at ($(P1) +(8.4,0)$) {NASDAQ detector};       
  \node[particle] (E2) at ($(P2)+(-2.3,0)$) {M2};

  \draw[dashed] (E1) -- (E2) node[midway, above] {$\vert \Phi^+\rangle$};

  \node[singlearrow] at ([yshift=1cm]P1) {};
  \node[singlearrow] at ([yshift=1cm]P2) {};
  \node at ([yshift=1.7cm]P1) {Correlation flipped?};
  \node at ([yshift=1.7cm]P2) {Correlation flipped?};

  \node[singlearrow] at ([yshift=-1cm]P1) {};
  \node[singlearrow] at ([yshift=-1cm]P2) {};
  \node at ([yshift=-1.7cm]P1) {Buy/Sell stock X};
  \node at ([yshift=-1.7cm]P2) {Buy/Sell stock Y};
\end{tikzpicture}
\caption{A physical implementation of a quantum strategy for the HFT scenario where the NYSE and NASDAQ servers are trading stock X and stock Y, respectively. Each server looks for a signal that the correlation between X and Y has flipped signs. This signal determines the measurement settings. M1 and M2 are two quantum memories at the different servers that are in the maximally entangled qubit state $\vert \Phi^+\rangle \coloneqq (\vert 00\rangle+\vert 11\rangle)/\sqrt{2}$, the state used in the optimal quantum strategy for the CHSH game. The corresponding measurement result determines whether to buy or sell. }
\label{fig:physical}
\end{figure}

Lastly, physical implementations of quantum strategies for this HFT scenario were also considered in~\cite{ding2024coordinating}. A schematic of such a physical implementation is shown in~\Cref{fig:physical}.

\noindent In particular, a heralded entanglement scheme~\cite{duan2001long} with a single pair of quantum memories that supports fast measurements \emph{is already sufficient} for the speeds involved for this NYSE-NASDAQ HFT scenario~\cite{ding2024coordinating}. That is, a quantum advantage in HFT can really be achieved with currently existing technologies. For a detailed overview of the physics and hardware in Bell inequality violation experiments, we refer the reader to~\cite{pan2012multiphoton}. 

\subsection{Distributed systems}
\label{subsec:distributed}
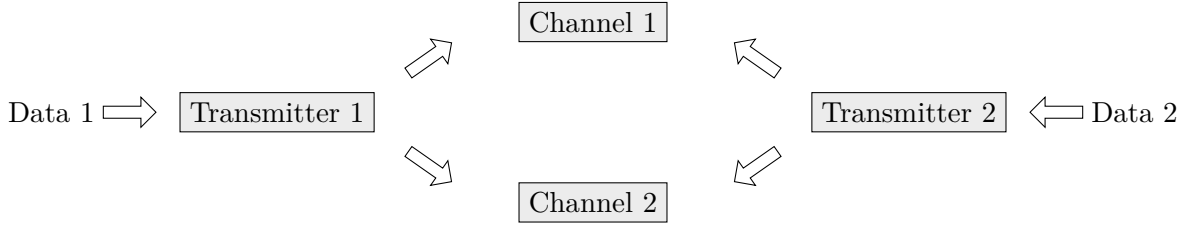
\begin{figure}[htbp!]
    \centering
    \begin{tikzpicture}[node distance=1.7cm,
  player/.style = {draw, rectangle, minimum width=.9cm, fill=gray!15},
  singlearrow/.style = {single arrow, draw, inner sep=0pt,
    minimum height=.7cm, minimum width=4mm, single arrow head extend=.3pt, rotate=-90},
  searrow/.style = {single arrow, draw, inner sep=0pt,
    minimum height=.7cm, minimum width=4mm, single arrow head extend=.3pt, rotate=-35},
    nearrow/.style = {single arrow, draw, inner sep=0pt,
    minimum height=.7cm, minimum width=4mm, single arrow head extend=.3pt, rotate=35},
    warrow/.style = {single arrow, draw, inner sep=0pt,
    minimum height=.7cm, minimum width=4mm, single arrow head extend=.3pt, rotate=180},
    earrow/.style = {single arrow, draw, inner sep=0pt,
    minimum height=.7cm, minimum width=4mm, single arrow head extend=.3pt, rotate=0},
    nwarrow/.style = {single arrow, draw, inner sep=0pt,
    minimum height=.7cm, minimum width=4mm, single arrow head extend=.3pt, rotate=145},
    swarrow/.style = {single arrow, draw, inner sep=0pt,
    minimum height=.7cm, minimum width=4mm, single arrow head extend=.3pt, rotate=-145}]
  
  \node[player] (P1) at (0,0) {Transmitter 1};
  \node[player] (P2) at ($(P1) +(8.4,0)$) {Transmitter 2};       
  \node[player] (C1) at (4.2,1.2) {Channel 1};
  \node[player] (C2) at (4.2,-1.2) {Channel 2};       


  \node[earrow] at ([xshift=-2cm]P1) {};
  \node[warrow] at ([xshift=2cm]P2) {};
  \node at ([xshift=-3cm]P1) {Data 1};
  \node at ([xshift=3cm]P2) {Data 2};

  \node[nearrow] at ([yshift=0.7cm,xshift=2cm]P1) {};
  \node[searrow] at ([yshift=-0.7cm,xshift=2cm]P1) {};
  \node[swarrow] at ([yshift=-0.7cm,xshift=-2cm]P2) {};
  \node[nwarrow] at ([yshift=0.7cm,xshift=-2cm]P2) {};
\end{tikzpicture}
\caption{The load balancing problem in ad hoc network routing. Multiple transmitters can send data via multiple channels. At any point each transmitter has a certain data rate. The total data rate in a single channel cannot exceed a certain threshold. We also want to minimize the number of channels used. }
\label{fig:qlb}
\end{figure}

Another domain where latency-constrained scenarios can occur is distributed systems~\cite{brandenburger2015quantum,hasanpour2017quantum,ding2024coordinating,da2025entanglement,arun2025faster,gardiner2026learning}. In particular, we consider systems in which networked protocols use randomization to make decisions in a decentralized manner. 
For example, multiple works have considered using quantum telepathy for \textbf{load balancing}. Load balancing is a concept that applies in routing for ad hoc networks as well as in assigning incoming customer requests to different backend servers. Load balancing for ad hoc networks, at a high level, is appropriately allocating the amount of data routed via different channels~\cite{hasanpour2017quantum}. This is illustrated in~\Cref{fig:qlb}.
\noindent The logic of the load balancing problem in ad hoc network routing is the following.
\begin{itemize}
    \item There are multiple transmitters that each need to transmit data at a certain rate. The data rate for transmitter $j$ is his input $i_j$.
    \item Each transmitter can choose from multiple channels to send their data. The channel chosen by transmitter $j$ is his output $o_j$.
    \item Each channel has a threshold combined data rate $r^*$ (channel capacity) that it can support.
    \item The total number of channels used should be minimized (to save energy or leave channels available for other purposes).
    \item Each transmitter does not know the data rates of other transmitters in real time due to communication latency.\footnote{Other possible reasons include the overhead of having centralized control as well as confidentiality (for instance, the transmitters could belong to competing telecommunication firms). Note that these reasons would make this into an isolated-party scenario. }
\end{itemize}
For simplicity, let the utility function $\mathcal U$ yield 1 if the transmitters choose channels such that the threshold is not exceeded for every channel and the number of channels is minimized,\footnote{The minimum number can be computed by solving the corresponding bin packing problem. } and let it yield 0 otherwise.
In the case of two transmitters, two channels, and two possible data rates $r, r'$ with equal probability for each transmitter, where $r<r'$, if 
\begin{align*}
    r+r' < r^* < 2r',
\end{align*}
then the load balancing problem exactly corresponds to the CHSH game. That is, when both transmitters are transmitting at rate $r'$, they should choose different channels. Otherwise, they should choose the same channel. In other words, we can achieve better load balancing in distributed systems using quantum entanglement. It is possible to obtain other nonlocal games with a quantum advantage by generalizing this to scenarios with more parties, inputs, and outputs.

\begin{figure}[htbp!]
    \centering
    \begin{tikzpicture}[node distance=1.7cm,
  player/.style = {draw, rectangle, minimum width=.9cm, fill=gray!15},
  particle/.style = {draw, circle, minimum size=.7cm, inner sep = 0.3, fill=gray!15},
  singlearrow/.style = {single arrow, draw, inner sep=0pt,
    minimum height=.7cm, minimum width=4mm, single arrow head extend=.3pt, rotate=-90}]
  
  \node[player] (P1) at (0,0) {Transmitter 1 detector};
  \node[player] (P2) at ($(P1) +(10.4,0)$) {Transmitter 2 detector};       
  \node[player] (Pn) at ($(P1)+(5.2,0)$) {Photon source}; 

  \draw[->,decorate, decoration={snake, amplitude=0.3mm, segment length=2mm}] (Pn) -- (P1) node[midway, above] {$\gamma$} ;
  \draw[->, decorate, decoration={snake, amplitude=0.3mm, segment length=2mm}] (Pn) -- (P2) node[midway, above] {$\gamma$} ;

  \node[singlearrow] at ([yshift=1cm]P1) {};
  \node[singlearrow] at ([yshift=1cm]P2) {};
  \node at ([yshift=1.7cm]P1) {Data rate 1};
  \node at ([yshift=1.7cm]P2) {Data rate 2};

  \node[singlearrow] at ([yshift=-1cm]P1) {};
  \node[singlearrow] at ([yshift=-1cm]P2) {};
  \node at ([yshift=-1.7cm]P1) {Choose channel 1 or 2};
  \node at ([yshift=-1.7cm]P2) {Choose channel 1 or 2};
\end{tikzpicture}
\caption{A physical implementation of a quantum strategy for the load balancing problem in ad hoc network routing. Each transmitter has a certain data rate. This rate determines the measurement settings. There is a photon source sending entangled photons to the two transmitters. The corresponding measurement result determines which channel to choose. }
\label{fig:physical2}
\end{figure}
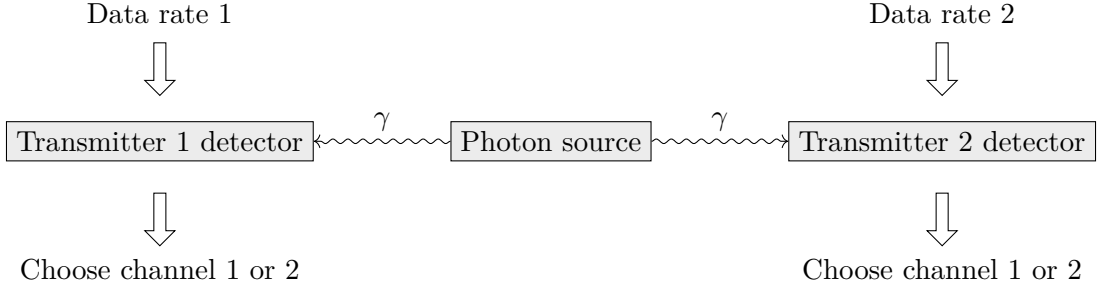 

\Cref{fig:physical2} shows a physical implementation of a quantum strategy for the load balancing problem in ad hoc network routing.
For distributed systems where different nodes are on the order of a hundred meters apart (the square root of the average area of a data center in 2024~\cite{techjury}) and are taking actions at microsecond timescales, a physical implementation of a quantum telepathy scheme may only need a standard MHz rate entangled photon source distributed via optical fiber without any quantum memories~\cite{ding2024coordinating}. Hence, the distance and timescales in distributed systems may be the perfect parameter regime for current hardware. To violate the CHSH inequality, distances of up to 10 km are also possible with such a physical implementation. For even larger distances, quantum memories may be needed so that photon loss does not preclude a quantum advantage~\cite{mermin1986new}.

\section{Isolated-Party Scenarios}
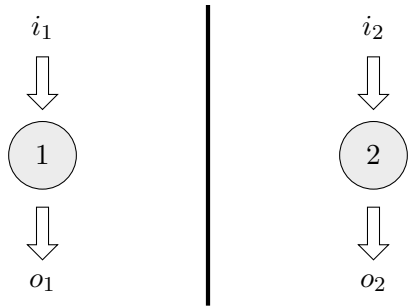
\begin{figure}[htbp!]
    \centering
    \begin{tikzpicture}[node distance=1.7cm,
  player/.style = {draw, circle, minimum width=.9cm, fill=gray!15},
  singlearrow/.style = {single arrow, draw, inner sep=0pt,
    minimum height=.7cm, minimum width=4mm, single arrow head extend=.3pt, rotate=-90}]
  
  \node[player] (P1) at (0,0) {$1$};
  \node[player] (P2) at ($(P1) +(4.4,0)$) {$2$};       

\draw[ultra thick] (2.2,2) -- (2.2,-2);

  \node[singlearrow] at ([yshift=1cm]P1) {};
  \node[singlearrow] at ([yshift=1cm]P2) {};
  \node at ([yshift=1.7cm]P1) {$i_{1}$};
  \node at ([yshift=1.7cm]P2) {$i_{2}$};

  \node[singlearrow] at ([yshift=-1cm]P1) {};
  \node[singlearrow] at ([yshift=-1cm]P2) {};
  \node at ([yshift=-1.7cm]P1) {$o_{1}$};
  \node at ([yshift=-1.7cm]P2) {$o_{2}$};
\end{tikzpicture}
\caption{An isolated-party scenario. There is an obstacle to communication, abstractly represented by the thick line in the middle. }
\label{fig:isolated}
\end{figure}

Non-communication in a real-world scenario may also be due to the parties being isolated from each other. That is, there is some \emph{obstacle to communication}. This could be due to a physical barrier between the parties, a lack of means to communicate, or a privacy concern~\cite{brandenburger2015quantum,szegedy2020systems,mironowicz2023entangled}. Such a situation is depicted in~\Cref{fig:isolated}.

\noindent Concrete examples include the following scenarios. A group of drones exploring an underwater cave as part of a scientific expedition may not be able to communicate because the cave walls are physical barriers to communication. Two cartographers, lost in a remote jungle without cellphone or satellite signal, cannot communicate because there are no available means of communication. Multiple nodes in an ad hoc network may not want to communicate their data rates since they belong to competing telecommunication firms (see~\Cref{subsec:distributed}).
Multiple parties attempting to communicate via radio in the presence of adversarial jammers need to choose a common frequency band but cannot communicate by construction, a scenario explored in a recent paper~\cite{wehner2026quantumadvantage}.
These examples are what we refer to as isolated-party scenarios.
We note that the framework in~\cite{ding2025quantum} can be also applied to isolated-party scenarios where the obstacle to communication only applies to a \emph{subset of parties}. 
We also mention that non-communication could be imposed by fiat. That is, an artificial rule is set so that the parties are not allowed to freely communicate. Examples include prisoners that are separately interrogated~\cite{jacobs2005entangled}, players in a game show~\cite{steane2000physicists}, and players in a card game~\cite{muhammad2014quantum,lin2020quantum}. However, for these examples, the rules were initially determined without taking into account the use of quantum resources. An appropriate modification of the rules can eliminate this quantum advantage.

Unfortunately, quantum telepathy for isolated-party scenarios may require much more advanced hardware than for latency-constrained scenarios. In particular, the obstacle to communication that engenders the isolated-party scenario \emph{may also be an obstacle to establishing entanglement between the parties}. The most common ways to establish entanglement with current technologies is either a centralized entangled photon source or entanglement swapping via an intermediate station. Such physical implementations provide the parties with a constant, fresh supply of entanglement. This is important because quantum systems suffer from decoherence, and so any newly established entanglement is usually short-lived. However, neither a centralized source nor an intermediate station may be viable in an isolated-party scenario. This implies that entanglement used in such a scenario would necessarily need to be long-lived. There are two possible approaches to this:
\begin{enumerate}
    \item Quantum error correction that can lengthen coherence times by many orders of magnitude.
    \item The discovery of a quantum system that is both extremely stable (very long coherence time) and can support high-fidelity quantum gates and measurements.
\end{enumerate} 
We do not expect such a quantum memory to be available in the near term. However, it is very easy to underestimate how fast quantum hardware can develop and improve. Furthermore, identifying applications of quantum technologies, even if they might lie far in the future, is essential to the field’s long-term prospects. Therefore, studying isolated-party scenarios remains an important research priority.

\subsection{Rendezvous problems}
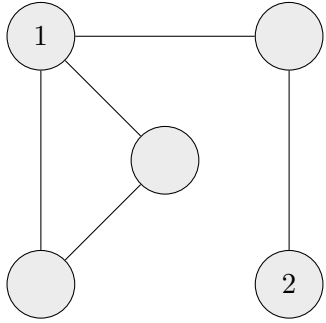
\begin{figure}[htbp!]
    \centering
    \begin{tikzpicture}[node distance=1.7cm,
  player/.style = {draw, circle, minimum width=.9cm, fill=gray!15},
  singlearrow/.style = {single arrow, draw, inner sep=0pt,
    minimum height=.7cm, minimum width=4mm, single arrow head extend=.3pt, rotate=-90}]
  
  \node[player] (V1) at (0,0) {$1$};
  \node[player] (V2) at ($(P1) +(3.3,0)$) {};       
  \node[player] (V3) at ($(P1) +(3.3,-3.3)$) {$2$};       
  \node[player] (V4) at ($(P1) +(0,-3.3)$) {};       
  \node[player] (V5) at ($(P1) +(1.65,-1.65)$) {};       

\draw (V1) -- (V2);
\draw (V1) -- (V5);
\draw (V1) -- (V4);
\draw (V4) -- (V5);
\draw (V2) -- (V3);


\end{tikzpicture}
\caption{A rendezvous problem. Party 1 is initialized on the northwest corner vertex while Party 2 is initialized on the southeast corner vertex. Their goal is to arrive at the same vertex simultaneously. }
\label{fig:rdv}
\end{figure}

Isolated-party scenarios are usually the context of \textbf{rendezvous problems}~\cite{alpern1995rendezvous}. This is a problem in classical computer science which involves a search on a graph. We draw the setup in~\Cref{fig:rdv}.
\noindent Each party is initialized on a vertex of a graph. At each time step, they are allowed to move to another vertex via an edge. The final goal is to rendezvous, that is, arrive at the same vertex at the same time. For example, for the graph and initial positions shown in~\Cref{fig:rdv}, if the two players move to the northeast corner vertex in the first step, then they successfully rendezvous. Such rendezvous problems arise in tasks such as finding a common channel of communication, drone navigation, or locating a stray traveler by a rescue team~\cite{mironowicz2023entangled}. In each case there is a corresponding obstacle to communication. 

A series of works~\cite{mironowicz2023entangled,viola2024quantum,tucker2024quantum} explored how to use quantum nonlocality to achieve a quantum advantage in rendezvous problems. Their approach is to model rendezvous problems by nonlocal games: let party $j$'s input $i_j$ be his initial vertex and his output $o_j$ be a sequence of edges which describes the path he will take. A time limit is set so that the path has a fixed length. For a wide variety of different graphs and variations on the rules of the game, a quantum advantage was identified. The quantum circuits that violate the corresponding Bell inequalities were even demonstrated on a quantum device in a laboratory setting~\cite{tucker2024quantum}.

\section{Discussion}
Quantum telepathy is a new form of quantum technology that 
\begin{enumerate}
    \item has a mathematically provable quantum advantage, and
    \item can be physically implemented with existing or near-term quantum hardware.
\end{enumerate}

Going forward, we need to identify more families of nonlocal games which manifestly model real-world problems. This would require a combination of theoretical as well as numerical research to explore the vast world of Bell inequalities. Beyond nonlocal games, latency-constrained (LC) games~\cite{ding2025quantum} is an even more promising direction to find applications. This is because in real-world problems, communication is often still possible, albeit among a subset of parties or with a latency cost. Furthermore, so far we have only mentioned cooperative games where multiple parties collaborate to maximize a global utility. We can also consider non-cooperative games where each party wants to maximize his own utility~\cite{auletta2021belief} and see if these can model real-world problems~\cite{khan2021quantum}. A quantum advantage would correspond to achieving a Nash equilibrium with higher social welfare than is possible with classical resources. 

It is also very important to experimentally demonstrate that the promised quantum advantage can be physically realized. Such experiments may prove more challenging than those that violate the CHSH inequality since they might require more complex quantum states and measurements. In particular, for scenarios involving more than two parties, multiparty entanglement may be necessary. Furthermore, the experiments need to be loophole-free~\cite{hensen2015loophole,shalm2015strong,giustina2015significant} in order for the quantum advantage to be genuine. For LC games, experimental demonstrations would also need real-time classical communication between parties near the speed of light~\cite{ding2025quantum}. Furthermore, for actual industrial applications, the nonlocal game (or LC game) should be defined using real-world data instead of a simple toy model. This may lead to a much more complicated experiment.

Lastly, we emphasize that quantum telepathy does not only concern real-world applications, but also fundamental physics~\cite{ding2025quantum}. As mentioned above, Bell inequalities and LC inequalities can be interpreted as fundamental limits on the correlations between classical parties when the latency constraint is comparable to the speed of light delay between different parties. Quantum entanglement can violate such inequalities. Already, the previous two sentences bring together the theory of relativity and quantum mechanics. As mentioned in~\cite{ding2025quantum}, we can also consider more complicated scenarios where the parties are in motion. If they move at relativistic speeds, defining the speed of light delay becomes more subtle. The parties could even be moving in curved spacetime due to a gravitational field. Deriving the corresponding fundamental limits on correlations for classical parties and seeing how they can be violated with quantum entanglement could be extremely interesting in its own right.

\paragraph{Acknowledgment} We would like to thank Rasoul Etesami, Qiming Wu, and Jiehang Zhang for helpful discussions.
DD would like to thank God for all of His provisions.

\bibliographystyle{unsrt}
\bibliography{ref}
\end{document}